# Analogous Viscosity Equations of Granular Powders Based on Eyring's Rate Process Theory and Free Volume Concept

Tian Hao
15905 Tanberry Dr., Chino Hills, CA 91705
Dated: August 18, 2015

**Abstract**

Granular powders can be successfully treated with kinetic theory and statistical mechanics that are typically applicable to thermal systems, though the granular powders are athermal systems and the conventional environmental temperature is too weak to drive particles to move. Once the granular temperature is analogously defined in line with that in thermodynamics, viscosity concept of thermal systems is naturally borrowed to describe the flowability of granular powders in this article. Eyring's rate process theory and free volume concept, which have been proved to be very powerful in dealing with many thermally activated phenomena in a wide variety of fields, are utilized to derive viscosity equations of granular powders under a simple shear. The obtained viscosity equations are examined only with empirical experimental observations in describing powder flowability, due to the lack of instruments and methodology for directly determining the viscosity of granular materials. The continuous shear thickening rather than the discontinuous shear thickening are predicted and found to be dependent on shear rate, the cohesive energy between particles, and the particle volume fraction, though the discontinuous shear thickening may still occur if certain conditions are met during shear, such as local particle volume fractions approach to the jamming point created by the shear induced inhomogeneity. A fundamental mechanism on how dry granular powders flow is proposed on the basis of what are demonstrated from the viscosity equations.The work presented in this article may lay a foundation to scale powder flowability in a more fundamental and consistent manner, at least providing an approach to consistently define the viscosity of granular powders. Since the same approaches are employed to derive the viscosity equations of granular powders as used to derive viscosity equations of liquids, colloidal suspensions, and polymeric materials, both athermal and thermal systems are thus unified with a single methodology.

## I. Introduction

Granular powders are widely recognized as athermal systems where the conventional environmental temperature is too weak to drive particles to move. However, there are many theoretical and experimental evidences indicate that granular powders could be successfully treated with the kinetic theory and statistic mechanics extracted from thermal systems [1,2,3,4,5,6,7,8,9,10] [11,12,13,14,15,16,17,18,19]. Under proper analogous definition of granular temperatures, thermodynamics is found to be a powerful tool for understanding the very common and important jamming phenomena in granular systems [18,19,20,21]. The jamming points can be analogously considered as the frozen phase transition points, where liquids loose the flowability when temperatures drop below the melting points and the whole systems turn into solids.

Flowability of granular powders is extremely important in powder handling processes in many industrial areas like food, pharmaceutical, mineral, and civil engineering, etc. [22,23,24]. Objectively describing the flowability of granular powders is relatively difficult due to the complexity of granular materials: Granular powders are usually non-continuum media and also compressible, thus the regular fluid mechanics that works for continuum and incompressible thermal fluids may not be applicable to granular powder systems [25,26]. Nonetheless, under a shear field externally applied or induced by the gravity of particles, the frictional rheology of granular powders can be uniformly described with that of liquids or colloidal suspensions [27,28], though the unification leans more on the empirical and phenomenological levels. The rheological profile similarities between liquids or colloidal suspensions and granular powders further demonstrate that both these two systems are fundamentally connected. If the temperature of granular powders can be properly defined in the way that it can attain same functionalities as the thermal environmental temperature, the viscosity and other rheological parameters of granular powders may be easily defined analogously for better describing powder flowability.

For granular powders of relatively fast moving particles, granular temperatures are usually defined with the kinetic energy connection between the temperature and the velocity, $\frac{3}{2} k_B T = \frac{1}{2} m v^2$, where $k_B$ is the Boltzmann constant, $m$ is the mass of the particle, and $v$ is the velocity of particles [29,30,31,32,33]. In this manner, the granular temperature remains a same original meaning as that in thermal systems and thus other thermodynamic principles may easily be applied to granular powders. In my previous work, both the granular temperature and the jamming temperature are defined with this method and the jamming phenomena are profoundly elucidated with new insights [21]. Those successes definitely drive me to believe that the analogous viscosity concept could be introduced to granular powders and thus the flowability of granular powders could be potentially described in more clear and fundamental levels.

For defining the viscosity of granular systems in analogy with that of thermal systems like liquids and colloidal suspensions, the analogous granular temperature instead of traditional temperature must be employed. As we already know, the viscosity concept of liquids and colloidal suspensions has been successfully used to describe the flow property and the viscosity can be accurately measured via proper viscometers or rheometers. However, the currently available dry powder rheometers or other flowability characterization instruments are unable to provide same level scientific data as we have experienced with rheometers/viscometers for liquids, colloidal suspensions, and solids [34]. The big challenge in dry granular powders area is that we cannot use conventional rheometers/viscometers to characterize granular powders, simply due to the non-continuum and compressible nature of granular materials. A more deep reason could be the disappearance of Brownian motions and the measured properties couldn't represent the statistically averaged macroscopic properties of whole systems. There is no a systematic way to define and measure viscosity of granular powders, as the traditional temperature and associated viscosity concepts are no longer working any more, and a new analogous temperature must be re-defined before the viscosity concept can be used for granular powders. The intrinsic athermal nature of granular materials requires a completely new framework for scaling the flowability. Although highly concentrated colloidal suspensions containing non-Brownian particles can be considered as athermal systems too as indicated in the literature [35]. However, such an athermal system is still different from granular powder athermal systems, as highly concentrated colloidal suspensions can still be characterized with regular rheometers with proper design of experiments [36], though the measured properties may not truly reflect the rheologial behaviors due to the athermal nature resulted from non-Brownian particles. Fundamentally, both granular powders and highly concentrated colloidal suspensions containing non-Brownian particles may be very similar, as the frictional contacts between particles, in addition to the geometrical confinement [37,38], should play a critical role in continuous and discontinuous shear thickening even jamming phenomena widely observed in both systems [28,35,39,40]. The puzzling issues in highly concentrated colloidal suspensions are that the discontinuous shear thickening phenomenon is not captured in the unified frictional rheology [28,39] and the predicted viscosity independence of shear rate is untrue [36,39]. As suggested in literature [39], the "absence of shear rate dependence hints a missing force (or time) scale in the description of dense suspensions", indicating that a new approach incorporating those parameters are necessary. The traditional temperature that is still used in highly concentrated athermal colloidal suspensions containing non-Brownian particles may reach its limitations and must be replaced with analogous granular temperature.

The viscosities of liquids including both pure and mixtures, colloidal suspensions, and polymeric materials with and without an external electric field have been extensively addressed in my previous publications [41,42], with the aid of Eyring's rate process theory [43] and free volume concept [44,45,46]. The obtained viscosity equations of liquids, colloidal suspensions, and polymeric solutions and melts, are consistent with experimental results. The Eyring's rate process theory has been proved to be very powerful in revealing physical mechanisms of chemical reactions, dielectric relaxations, resonance energy transfer and many other thermally activated motions [47]. The free volume concept has been widely used to determine various equilibrium properties of both solid and liquids [48,49]. The popular empirical tap density equations of granular powders, the logarithmic and stretched exponential laws, are successfully derived with the Eyring's rate process theory and free volume concept [50,51]. All those evidences suggest that the Eyring's rate process theory and free volume concept are very useful tools in dealing with both thermal and athermal systems. Therefore, the Eyring rate process theory and the free volume concept will be employed again for defining the analogous viscosity of granular powders in this article. The success of this attempt will potentially lead to the unification of obtaining viscosity equations across both thermal and athermal systems from liquids, colloidal suspensions, polymeric materials, to granular powders with a single methodology.

The article is arranged as follows: First, the viscosity equations of granular powders will be analogously derived only with Eyring's rate process theory. In next section, Eyring's free volume concept will be introduced in and the viscosity equations will be derived with both the Eyring's rate process theory and Eyring's free volume concept. The free volume calculated with my own method will be introduced from my previous article afterwards, and the viscosity equations of granular powders will be derived with the Eyring's rate process theory and my own free volume calculation. All those newly obtained viscosity equations will be compared with with empirical experimental observations. Since there is no a method available for experimentally measuring the viscosity of granular powders, direct even semi-quantitative comparison with experimental results is unfortunately not provided. The discussions and conclusions will be given in the end. This is just an initial step of introducing analogous viscosity to granular powders based on analogous granular temperature, the focus will thus be placed on the viscosity of granular systems for simplicity reason and other rheological concepts will be minimized.

## II. Theory
### 1. Viscosity equation derived directly from Eyring's rate theory

Flow as a rate process was proposed by Eyring [52] and summarized in the book [43]. Eyring's theory is originally for molecular systems, and is proved that this rate process theory can be used to predict the viscosity of colloidal suspensions and polymeric materials, too [41,42]. We may move one step further to use this theory for granular powders, for the reason that the colloidal suspensions are the particles dispersed in a continuous liquid medium, while the granular systems are the particles dispersed in continuous air medium, if an analogy really has to be made between those two systems. Consider two layers of particles in a powder bed have a distance $d$ apart and relative velocity between those two layers is $v$, see Figure 1. Under a shear stress $f$, the

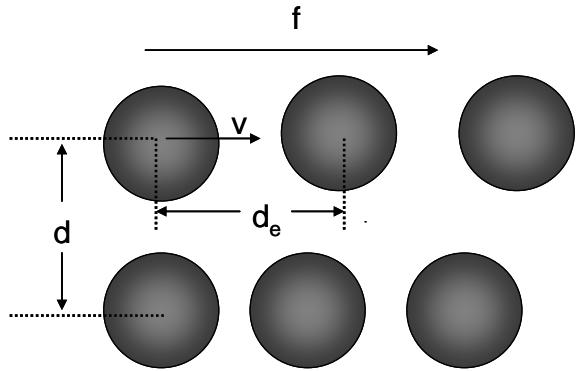

*Figure 1. Distance between two layers of particles in a powder bed under a shear force f. $d_e$ is the distance between two equilibrium positions under a shearing flow.*

viscosity, $\eta$, according to the definition, can be expressed as

$$\eta = \frac{f}{v/d} = \frac{fd}{v} \tag{1}$$

In analogy with Eyring's rate process theory, one may assume that a particle moves from one equilibrium position to another needs to overcome the energy barrier, the activation energy, $E_0$, as shown in Figure 2. The applied shearing force will reduce the height of the energy barrier in the flowing direction by $\Delta E$, while it will raise the height of the energy barrier in the opposite direction by the same amount. The number of times a particle passes over the barrier per second may be given by [43]:

$$k = \frac{k_B T_{gp}}{h} \bullet \frac{F_a}{F_i} e^{-E_0/k_B T_{gp}} \tag{2}$$

where $T_{gp}$ is granulotemperature, $k_B$ is Boltzmann constant, $h$ is Planck constant, $F_a$ and $F_i$ are the partition functions for unit volume of the particles in activated and initial states, $E_0$ is the activation energy. In the original treatment, $T_{gp}$ in Eq. (2) is the conventional temperature $T$. $T_{gp}$ is defined in the way that it has the conventional temperature functionalities and thus can replace $T$ in an analogous manner. Since the energy barrier of the flowing direction is lowered, so the specific rate in the flowing direction, $k_f$

$$k_f = \frac{k_B T_{gp}}{h} \bullet \frac{F_a}{F_i} e^{-(E_0-\Delta E)/k_B T_{gp}} = k e^{\Delta E/k_B T_{gp}} \tag{3}$$

The specific rate in the backward direction $k_b$

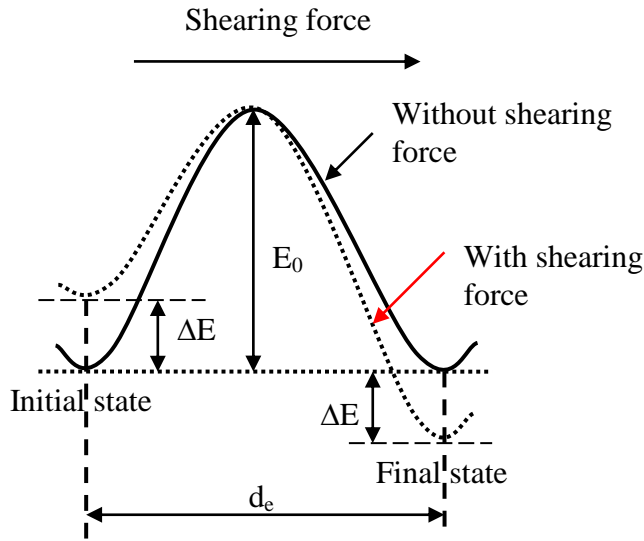

Figure 2. Potential energy barrier for viscous flow with and without a shearing force. From Hao, T., The electrorheological fluids: Non-aqueous suspensions, 2005, Amsterdam: Elsevier.

$$k_b = k e^{-\Delta E/k_B T_{gp}} \tag{4}$$

The net rate in the flowing direction ($k_f$ - $k_b$) times by the distance between two equilibrium positions $d_e$ should be the velocity of flowing

$$v = d_e(k_f - k_b) = d_e k(e^{\Delta E/k_B T_{gp}} - e^{-\Delta E/k_B T_{gp}})$$
$$= 2d_e k \sinh \Delta E / k_B T_{gp} \tag{5}$$

Combining Eq.(1) with Eq.(5) yields

$$\eta = \frac{fd}{2d_e k \sinh \Delta E / k_B T_{gp}} \tag{6}$$

Both $d$ and $d_e$ have the particle size dimension, and $\Delta E/(k_B T_{gp}) << 1$ in a granular system. Eq. (6) may be simplified as ($d \approx d_e$, $\sinh x \approx x$ for small x)

$$\eta = \frac{fk_B T_{gp}}{2k\Delta E} = \frac{hf}{2\Delta E} \bullet \frac{F_i}{F_a} e^{E_0/k_B T_{gp}} \tag{7}$$

Suppose that the particle moves from one equilibrium position to another is a kind of rate controlled reaction, the equilibrium constant $k_e$

$$k_e = \frac{F_a}{F_i} e^{-E_0/k_B T_{gp}} \tag{8}$$

and the thermodynamic relationship between the equilibrium constant and the Gibbs free energy

$$k_e = e^{-\Delta G/RT_{gp}} \tag{9}$$

$\Delta G$ is the standard Gibbs free energy, and R is gas constant. Thus Eq. (7) is therefore rewritten as:

$$\eta = \frac{hf}{2\Delta E} e^{\Delta G/RT_{gp}} = (\frac{hf}{2\Delta E} e^{-\Delta S/R}) e^{\Delta H/RT_{gp}} \tag{10}$$

$\Delta H$ is the enthalpy and $\Delta S$ is entropy. Since $\Delta E$ is dependent on $f$, $f/\Delta E$ thus can be taken as a constant. $\Delta S$ can also be taken as a constant under the assumption that the molar volume of a granular system does not largely change with granulotemperature. Eq. (10) may be written as:

$$\eta = Ae^{E/RT_{gp}} \tag{11}$$

A is a constant. Eq. (11) is the commonly used Arrhenius equation [53]. From Eq. (11), one may obtain several viscosity equations after replacing the granulotemperature $T_{gp}$ defined in my another article [21]. For a granular powder under a simple shear, the granulotemperature may be expressed as [21]:

$$T_{gp} = \frac{8}{9} \frac{\pi \sigma r^3 \dot{\gamma} \rho_t t}{k_B \rho_b} \tag{12}$$

where $r$ is the particle radius, $\sigma$ is the shear stress, $\dot{\gamma}$ is the shear rate, $\rho_t$ and $\rho_b$ are the true and bulk density of the granular powder, $t$ is time, and $k_B$ is the Boltzmann constant. Substituting Eq. (12) into Eq. (11), one may obtain the viscosity of a granular powder under a simple shear:

$$\eta = Ae^{\frac{9E\rho_b}{8N_A\pi\sigma r^3 \dot{\gamma}\rho_t t}} \tag{13}$$

where $N_A = 6.02 \times 10^{23} mol^{-1}$, the Avogadro's number. Eq. (13) indicates that the viscosity of a granular powder under a simple shear is dependent on the shear stress applied to the powder, shear rate created in the system, the true and bulk densities, and the sizes of particles. It is much more sensitive to the size of particles: the logarithm viscosity is inversely proportional to the cube of particle radius. In other words, anything that can induce particle size change, such as the humidity and electrostatic charge that control particle aggregation or agglomeration, will dramatically change the viscosity of a granular system. For granular powders under a vibration or on a slope, one may easily obtain the viscosity equations via simply substituting the correspondent granulotemperatures from my other article [21] into Eq. (11). The viscosities under a vibration or on a slope are obviously different from that of same powders under a simple shear due to the different granulotemperatures, further indicating that granular powders don't have a fixed viscosity. The measured viscosity should change with the applied shear stresses and shear rates.

2. **Viscosity equation derived from Eyring's rate theory and Eyring's free volume estimation.**

As mentioned earlier, the free volume concept was introduced by Eyring [52] in dealing with the viscosity of liquids in analogy with gases that can be assumed to consist of molecules made up of "holes" moving about in matter. The free volume is the space unoccupied by molecules and is probably smaller than that occupied by the molecules. It is reasonable to assume that there are a lot of holes in a liquid and the free volume $V_f$ of a liquid molecule is related to the molar volume of the liquid $V$, the molecules packing contant c, and the energy of vaporization per molecule per mole, $\Delta E_{vap}$. According to Eyring, the free volume may be expressed as [43]:

$$V_f = \left(\frac{cRT}{\Delta E_{vap}}\right)^3 \frac{V}{N_A} \tag{14}$$

$N_A$ is the Avogadro number. Note that the gas constant $R = k_B N_A$, so $RT$ is the molar thermal energy in a liquid. The constant c is the number of molecules in a packing unit cell, thus the term $cRT/\Delta E_{vap}$ scales the energy needed to separate the molecules per unit cell in relative to the energy needed to bind those molecules together. Since $V$ is the molar volume of a liquid, $V/N_A$ may be considered as the average volume per molecule and $V_f^{1/3}$ thus may scale how difficult to separate the molecules per unit cell in a distance of $(V/N_A)^{1/3}$. Based on above analysis, one would rather name $V_f$ as the "energy" needed to create a "holes" in a liquid, instead of its original name, free volume. The partition function of a molecule of liquid may be expressed as [46]:

$$F_i = \frac{(2\pi m k_B T)^{3/2}}{h^3} V_f b_l e^{-\Delta E_{vap}/RT} \tag{15}$$

where $b_l$ is the combined vibrational and rotational contribution, and $m$ is the mass of a liquid molecule. Since at the activated state a molecule has one degree of translational freedom less than that at the initial state [43]:

$$\frac{F_i}{F_a} = \frac{(2\pi m k_B T)^{1/2}}{h} V_f^{1/3} \tag{16}$$

Note that in the original definition, $F_i$ and $F_a$ are partition functions at initial and activated states per unit volume. However, $V_f$ is merely the free volume of an individual molecule. The total free volume per unit volume should be the number of molecules per unit volume multiplied by the free volume of a molecule. Since we concentrate on the ratio between $F_i$ and $F_a$, the factor of the number of molecules are omitted from Eq. (15) and (16). Substituting Eq. (14) and (16) into Eq. (7) and replacing $T$ with $T_{gp}$ yields

$$\eta = \frac{hf}{2\Delta E} \cdot \frac{(2\pi m k_B T_{gp})^{1/2}}{h} V_f^{1/3} e^{E_0/k_B T_{gp}}$$
$$= \frac{fcRT_{gp}}{2\Delta E_{vap}\Delta E} \left(\frac{V}{N_A}\right)^{1/3} (2\pi m k_B T_{gp})^{1/2} e^{E_0/k_B T_{gp}} \tag{17}$$

Eq. (17) seems to be reasonable in term of the relationship with $V_f^{1/3}$, as the viscosity should proportionally increase with the "energy" needed to create "holes" or so-called free volume. $V_f^{1/3}$ is apparently related to the "free distance" that a particle can move freely in the system. Intuitively, the viscosity should decrease with the increase of the free volume. In other words, a system of a larger free volume should have a lower viscosity as molecules may easily move to one equilibrium position to another under a shear force. The shear stress $f$ multiplied by the molar volume $V$ and then divided by Avogadro number $N_A$ should be the energy applied to each molecular, $\Delta E$

$$fV/N_A = \Delta E,$$
$$\frac{f}{\Delta E} = \frac{N_A}{V} \tag{18}$$

Substituting Eq. (18) into Eq. (17) leads

$$\eta = \frac{cR}{2\Delta E_{vap}} \left(\frac{N_A}{V}\right)^{2/3} (2\pi m k_B)^{1/2} T_{gp}^{3/2} e^{E_0/k_B T_{gp}} \tag{19}$$

Eq. (19) shows how the viscosity of a granular powder changes with the granulotemperature and other physical parameters. It also indicates that the constant $A$ in Eq. (13) is dependent on granulotemperature, too. Taking the granular powder under a simple shear as an example, one may obtain the viscosity of a granular powder by simply substituting $T_{gp}$ with Eq. (12) and $m$ with $\frac{4}{3}\pi r^3 \rho_t$

$$\eta = \frac{c\pi^2 N_A}{2\Delta E_{vap}} \left(\frac{N_A}{V}\right)^{2/3} \left(\frac{4}{3}\pi r^3 \rho_t\right)^{1/2} \left(\frac{8\sigma r^3 \dot{\gamma} \rho_t t}{9\rho_b}\right)^{3/2} e^{\frac{9E_0 \rho_b}{8\pi\sigma r^3 \dot{\gamma} \rho_t t}} \qquad (20)$$

$V$ is the mole volume of particles and thus may be represented as:

$$V = \frac{4}{3}\pi r^3 N_A \frac{\rho_t}{\rho_b} \qquad (21)$$

Substituting Eq. (21) into Eq. (20) yields

$$\eta = \frac{cN_A}{3\Delta E_{vap}} \left(\frac{4\pi}{3}\right)^{11/6} \rho_t^{1/2} \left(\frac{\rho_t}{\rho_b}\right)^{5/6} r^4 \left(\sigma\dot{\gamma} t\right)^{3/2} e^{\frac{9E_0 \rho_b}{8\pi\sigma r^3 \dot{\gamma} \rho_t t}} \qquad (22)$$

where c is a constant dependent on how particles pack in the system and equals to 2 for the cubic packing structure. Originally, $\Delta E_{vap}$ is the molar vaporization energy required for a liquid transferred into a gas, and may be considered as the energy that overcomes the binding interactions between molecules in liquid states. In a granular system, $\Delta E_{vap}$ may be considered as the energy required for separating particles to such a large distance that there is no interaction force between particles, i.e., the particles reach the granular gas state. The cohesive forces responsible for the interactions between particles should be directly related to the vaporization energy. Since the energy of activation for the viscous flow is related to the work required to form holes in the liquid, the activation energy $E_0$ may be expected to be some fraction of the evaporation energy [43]:

$$E_0 N_A = \frac{\Delta E_{vap}}{n} \qquad (23)$$

where *n* is a number factor that gives an indication of the size of the holes necessary for a viscous flow and is of the values from 2 to 5 [43]. Note that $E_0$ is the activation energy required for a molecule and $\Delta E_{vap}$ is the molar vaporization energy, thus the Avogadro constant $N_A$ is needed to reflect the difference between $E_0$ and $\Delta E_{vap}$ in the calculation. According to Eyring, for most liquids of symmetrical molecular structure like water, *n=2.45*, nearer 3, and for polar molecules with non-spherical symmetry like long-chains hydrocarbons, *n* is about 4. For simplicity reason, one may take *n=3* for granular materials. Under those assumptions and renaming $\Delta E_{vap}$ as $\Delta E_{coh}$, the cohesive energy between particles, Eq. (22) may be rewritten as:

$$\eta = \frac{cN_A}{3\Delta E_{coh}} \left(\frac{4\pi}{3}\right)^{11/6} \rho_t^{1/2} \left(\frac{\rho_t}{\rho_b}\right)^{5/6} r^4 \left(\sigma\dot{\gamma} t\right)^{3/2} e^{\frac{3\Delta E_{coh}\rho_b}{8\pi\sigma r^3 \dot{\gamma} \rho_t t N_A}} \qquad (24)$$

Eq. (24) clearly shows that viscosity of a granular powder is a function of particle size, true and bulk densities, packing structure, cohesive forces scaled with the separation energy, the applied shear stress, shear rate, and amazingly the time. Qualitatively, this equation makes sense, as powder flowability is empirically found to be complicated and is dependent on those properties.

This equation will be evaluated in more details below for acquiring some basic ideals on how those parameters will change the analogous viscosity in granular powders.

First, let's acquire some ideas on the scale of the cohesive energy between particles, $\Delta E_{coh}/N_A$, based on the experimental results. The direct Atomic Force Microscopy (AFM) measurements of interparticle forces indicate that the median cohesive and/or adhesive forces between Paracetamol and common pharmaceutical excipients like Povidone, Crospovidone, Pregelatinized Starch, Stearic Acid, and Magnesium Stearate ranges 1.2 to 50 $nN$ [54]. Suppose that under a shear force particles may move in a distance of the order of the particle size, 100 μm, then the cohesive energy between a pair of particles ranges from $1.2 \times 10^{-13}$ ~ $50 \times 10^{-13}$ J, which scales the cohesive energy per particle $\Delta E_{coh}/N_A$. According to Eq. (24), this estimation may be reasonable as $\Delta E_{coh}/N_A$ should be quite close to the value expressed in the following equation:

$$\Delta E_{coh}/N_A \approx \left( \frac{3\rho_b}{8\pi\sigma r^3 \dot{\gamma} \rho_t t} \right)^{-1} \tag{25}$$

Otherwise, the term $e^{\frac{3\Delta E_{vap}\rho_b}{8\pi\sigma r^3 \dot{\gamma} \rho_t t N_A}}$ could become too large and Eq. (24) becomes invalid. Taking 100 μm particles as an example and assuming that $\sigma \dot{\gamma} t = 0.1, \rho_b/\rho_t = 0.2$, the typical values for a powder system, thus $\Delta E_{vap}/N_A = 41.87 \times 10^{-13}$ J, in the range of the cohesive energy obtained with AFM measurement. Once gaining an idea on the ranges of cohesive energy between particles, one may easily compute the viscosity of a granular powder with Eq. (24) against $\Delta E_{vap}/N_A$, which is shown in Figure 3, under the conditions of $c=2$, $r=100$ μm, $\sigma \dot{\gamma} t = 0.1, \rho_b/\rho_t = 0.2$. It is interesting to see that the viscosity doesn't monotonically change with the cohesive energy. Instead, there is a minimum viscosity occurring at $\Delta E_{vap}/N_A = 5\times 10^{-12} J$. When the cohesive energy is very small, the viscosity is surprisingly high, seemingly contradicting to commonsenses but actually reasonable as the particles are hard to move with an external shearing field due to the lack of "stickiness" between particles. In other words, a sticky particle moving under a shearing field could induce another particle to move together if the cohesive force between them is strong enough. This is the reason that the viscosity decreases with the further increase of cohesive force, as it is relatively easier for particles to move together as those particles are cohesively bonded. This trend continues to the point where the cohesive force is so strong that it starts to compete with the shearing force. At such cases the viscosity increases with further increase of the cohesive forces, as the cohesive forces between particles create a resistant force to the shearing force. In contrast to the slow viscosity decrease before the minimum turning point, the viscosity abruptly increases once the cohesive forces exceed $\Delta E_{vap}/N_A = 5\times 10^{-12} J$. Note that the predicted viscosity at the minimum point is only over two times higher than the viscosity of water at 20°C, $1\times 10^{-3}$ Pa.s. As mentioned earlier the typical cohesive and/or adhesive energy of drug and commonly-used excipients measured with AFM is in the range $1.2\times 10^{-13} \sim 50\times 10^{-13}$ J, the corresponded viscosity predicted with Eq. (24) is about $3.6\times 10^{-2} \sim 2.32\times 10^{-3} Pa.s$, which seems to be consistent with experience of pouring and stirring those powders when comparing with water.

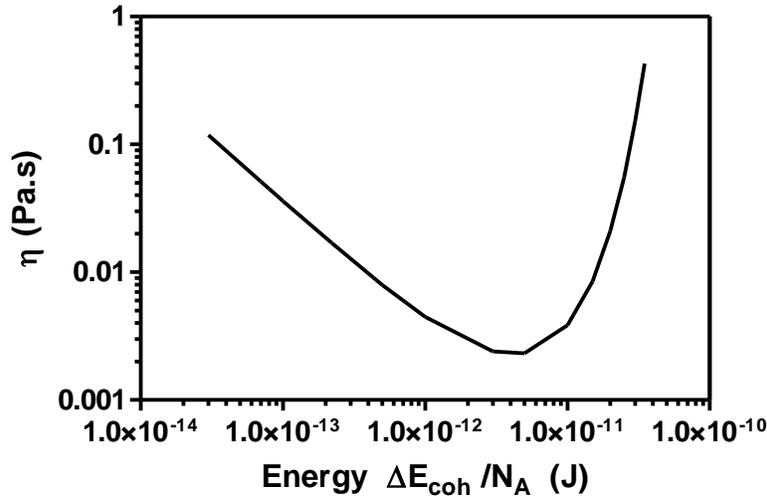

*Figure 3 The viscosity vs. the cohesive (or adhesive) energy between particles predicted with Eq. (24). Particle radius $r=100$ μm, $c=2$, $\sigma \dot{\gamma} t = 0.1, \rho_b/\rho_t = 0.2$*

It would be curious to see how the viscosity is going to change with the shear rates. The viscosity predicted with Eq. (24) vs. the shear rates under conditions $r=100$ μm, $\rho_t = 1000 kg/m^3, \rho_t/\rho_b = 5, \Delta E_{vap}/N_A = 5\times 10^{-13} J, \sigma t = 1 Pa\bullet s$, is shown in Figure 4. The predicted flowing profile is quite similar to the shear-thickening phenomenon observed in concentrated aqueous suspensions of solid particles [55,56]. In all known cases of shear-thickening phenomena observed in aqueous colloidal suspensions, there is a region of shear-thinning at lower shear rates [57], which corresponds to the region below the shear rate of 0.01s$^{-1}$ in Figure 4. In literature [58,59], the shear thinning and shear thickening phenomena could be explained with the following physical picture: (a) the formation of sliding layers; (b) the breakdown of these layers and large enough aggregates formed by the fragments of those broken layers as shear rate increases; The shear thinning phenomenon can be readily attributed to the slipping of layers passing each other, while the shear-thickening phenomenon is resulted from the breakdown of layers and formation of large aggregates at high shear rates; There is a critical shear rate at which the shear thinning gives away and shear-thickening starts to occur as the shear rate continues to increase beyond this critical point; If the average dimensions of shear-induced aggregates are large enough in flow-gradient direction to jam the flowing gap, an abrupt shear-thickening would occur, typically called colloidal jamming phenomenon [58,59,60,61,62,63], or called the discontinuous shear thickening [38]; Jamming means that an amorphous system develop a yield stress and it could be unjammed if the applied shear stress is strong enough. The granular powders behave exactly same as concentrated colloidal suspensions in term of how viscosity changes with shear rates: At low shear rates granular systems display a shear thinning phenomenon; there is also a critical shear rate at which the shear thinning gives away to the shear thickening. The critical shear rate at current calculation conditions is $8\times10^{-3}$ s$^{-1}$, a relatively low shear rate in comparison with colloidal suspensions [55,57]. As the shear rate increases further to the point where the generated shear stress ($\eta\dot{\gamma}$) is larger than the applied shear stress σ, the granular systems become "jammed", which is marked in Figure 4. The only different feature from colloidal jamming is that the viscosity of jammed granular systems doesn't abruptly increases as shear rate increases further. In other words, in granular powders the jammed states may be easily melt under a higher shear rate and then jam again, relatively slowly pushing the viscosity higher and higher, which is addressed in detail in my another article [21].

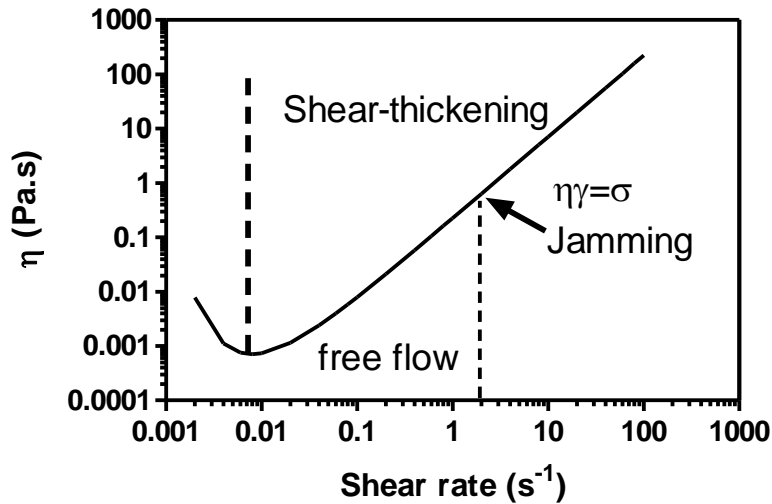

Figure 4. The predicted viscosity with Eq. (24) vs. shear rate under conditions r=100 μm, $\rho_t = 1000 kg/m^3, \rho_t/\rho_b = 5, \Delta E_{vap}/N_A = 5\times 10^{-13} J, \sigma t = 1$ Pa•s.

It is worth mentioning that the shear thickening predicted with Eq. (24) and demonstrated in Figure 4 is dependent of shear rate. As indicating in the introduction section, non-Brownian hard-sphere suspensions are expected to have viscosity independent of shear rate and shouldn't show the discontinuous shear thickening phenomena, due to the disappearance of the Brownian motions. However, the experimental results just give opposite evidences. The discontinuous shear thickening phenomena are not captured with Eq. (24) either, but the shear rate dependence is predicted.

The jamming phenomenon in granular systems has been extensively addressed in literature. The common assertions both theoretically and experimentally is that the jamming is a true second-order critical phenomenon [64,65], in analogy with the glass-transition in molecular systems [66]. A thermodynamic unification of jamming phenomena observed in both granular systems and molecular systems of glass transition was claimed to be established [32]. Unlike molecular glasses where increasing temperatures would melt the glasses and make the systems unjammed, increasing granulotempreature but keeping the applied shear stress same would cause the granular systems jammed. This apparent paradox may come from the way that the calculation is done: keeping shear stress unchanged but simply increasing the shear rates. Practically, one may be unable to freely increase shear rate if the applied shear stress is too small. This may only happen if the shear stress is strong enough to yield any resistant forces from the granular materials during shearing. Nevertheless, if the granulotemperature increase is resulted from the applied shear stress increase, the jammed granular systems would definitely melt, or called unjammed. For molecular glasses like polymer glasses, both heating (increase temperature) and strong shear stress may melt glasses [67,68], while for granular systems, those two approaches of unjamming methods merge into one rout, as the granulotemperature is defined to linearly increase with applied shear stress, see Eq. (12).

The applied shear stress plays an exact same role as shear rate in the viscosity equation, Eq. (24), and one may obtain a same viscosity profile against shear stress as shown in Figure 4 for viscosity against shear rates. The viscosity calculated with Eq. (24) against shear stress is shown in Figure 5, under conditions $r=100$ μm, $\rho_t = 1000 kg/m^3, \rho_t/\rho_b = 5, \dot{\gamma} t = 1$ for two cohesive energies, $\Delta E_{vap}/N_A = 5 \times 10^{-13} J$, and $\Delta E_{vap}/N_A = 9 \times 10^{-13} J$. There is a very steep decrease of viscosity at very low shear stresses, and there is no a plateau before that, implying that granular materials don't have a yield stress, different from concentrated colloidal suspensions [59,69,70]. At very low shear stresses the viscosity of granular materials is about $10^7$ Pa.s, quite close to that of Bitumen, $10^8$ Pa.s. However, the viscosity steeply decreases even a very small shear stress is applied on the system, signaling that granular materials seem to be very "soild" but actually very weak. As the shear stress continues to increase the viscosity decreases until reaches a minimum point. Beyond this critical point, the viscosity would increases with the applied shear stresses, indicating that a strong shear-thickening phenomenon occurs in this region. Neither Bingham yield stress model [71] nor its modified forms like Herschel-Bulkey [72] and Casson equations [73] can describe the relationship between viscosity and shear stress after the critical shear stress point, an apparent yield-dilatant behavior ubiquitously spotted in many granular systems [74,75,76,77]. An interesting phenomenon is that when one system of higher cohesive energy than another, its viscosity at very low shear stresses within the shear thinning region is higher than that of the low cohesive energy system; however, the viscosity becomes lower in shear-thickening region once shear stress is beyond the critical point. This may be because the dilatancy is resulted from the easy rearrangement of particles. A strong cohesive force between particles may increase difficulty for particles to re-arrange the configurational structures. Note that for a granular system with a higher cohesive energy the critical shear stress point shifts to a higher value, implying that a higher shear stress is needed to fully break down the particle structures in the powder bed.

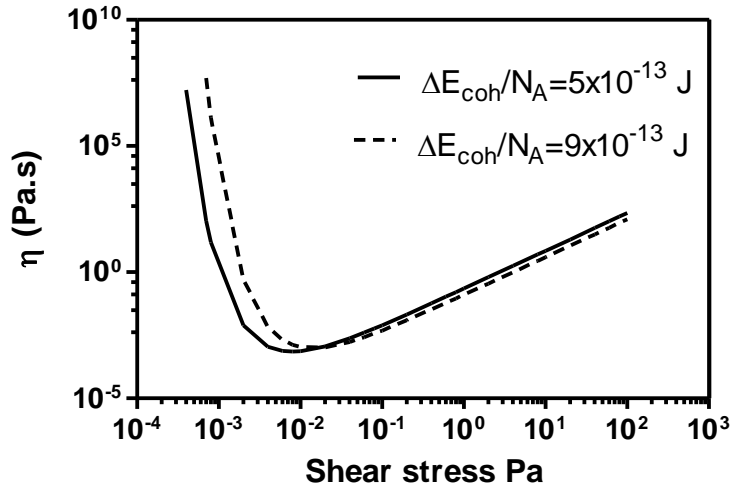

*Figure 5. The predicted viscosity with Eq. (24) vs. shear stress calculated under conditions $r=100$ μm, $\rho_t = 1000 kg/m^3, \rho_t/\rho_b = 5, \dot{\gamma} t = 1$ for two cohesive energies $\Delta E_{coh}/N_A = 5 \times 10^{-13} J$ and $\Delta E_{coh}/N_A = 9 \times 10^{-13} J$.*

It would be very insightful to see how the viscosity is going to change with the true density of granular powders based on Eq. (24), as we empirically know that metallic particles usually flow much better than polymeric particles due to the true density difference. This topic is extensively discussed in my other article [78]. For avoiding the redundancy, only the main results are presented here and please refer the literature [78] for detailed information. The predicted viscosity with Eq. (24) sharply decreases with the true density varying from about one to three, in line with empirical observations: Organic or polymeric materials of a true density about 1 g/cm$^3$ typically have a poor flowability in comparison with metallic particles of a true density above 6 g/cm$^3$. Even the interparticle cohesive energy between particles and particle sizes are same for both organic or polymeric and metallic particles, metallic particles are predicted to have a lower viscosity, thus flowing much better than low density particles. Another interesting prediction is that particles of higher cohesive energy give larger viscosity initially at low density regions, but show lower viscosity at high density regions. This seems to be intuitive and reasonable, as particle with stronger interaction forces are hard to move at the beginning; Once they start to flow, particles may flow together collectively due to strong cohesive forces, showing lower viscosities. More importantly, Eq. (24) provides theoretical bases for two most popularly used empirical flowability indicators [54,79], Carr index [80] and Hausner ratio [81], which are defined as:

$$CI = \frac{\rho_{tap} - \rho_b}{\rho_{tap}}$$

$$H_r = \frac{\rho_{tap}}{\rho_b}$$

(26)

where $CI$ is Carr index and $H_r$ is Hausner ratio, $\rho_{tap}$ is the ultimate (equilibrium) tap density that will not change with the further increase of tap numbers. Obviously, $CI = 1 - 1/H_r$, and $\rho_{tap}$ should be smaller than but quite close to the true density, $\rho_{tap} = \Phi_m \rho_t$, where $\Phi_m$ is the maximum packing fraction at the equilibrium tapping state. One may reach:

$$\frac{\rho_{tap}}{\rho_b} = \Phi_m \frac{\rho_t}{\rho_b}$$

(27)

Empirically, a Carr's Index greater than 0.18 ($H_r$ is about 1.22) is considered to be an indication of poor flowability, and below 0.15 ($H_r$ is about 1.18) of good flowability [82,83], though many experimental evidences substantiate that Hausner ratio and Carr index are quite scarce and inaccurate flowability indicators [84]. This criterion implies that high Hausner ratio leads to a high viscosity, which is obvious on the basis of Eq. (24), though Eq. (24) provides a much more complicated relationship between the viscosity of powders and the Carr index or Hausner ratio. Most pharmaceutical active ingredients and excipients are either small molecular organic materials or polymers, and the true densities of those materials are quite similar and close to 1 g/cm$^3$. It would be reasonable to assume that the true density is a constant. Differentiating the viscosity expressed in Eq. (24) against $(\rho_t/\rho_b)$ under an assumption that $\rho_t$ is a constant yields:

$$\frac{d\eta}{d(\rho_t/\rho_b)} = \frac{cN_A}{3\Delta E_{vap}} \left(\frac{4\pi}{3}\right)^{11/6} \rho_t^{1/2} r^4 \left(\sigma \dot{\gamma} t\right)^{3/2} e^{\frac{3\Delta E_{coh}\rho_b}{8\pi\sigma r^3 \dot{\gamma} \rho_t t N_A}}$$
$$\left[\frac{5}{6}(\frac{\rho_t}{\rho_b})^{-1/6} - (\frac{\rho_t}{\rho_b})^{-7/6}\left(\frac{3\Delta E_{coh}}{8\pi\sigma r^3 \dot{\gamma} t N_A}\right)\right] \quad (28)$$

Suppose $\frac{d\eta}{d(\rho_t/\rho_b)} = 0$, one may readily obtain

$$\frac{\rho_t}{\rho_b} = 1.2 \times \frac{3\Delta E_{coh}}{8\pi\sigma r^3 \dot{\gamma} t N_A} \quad (29)$$

or

$$\frac{\rho_{tap}}{\rho_b} = 1.2 \times \frac{3\Delta E_{coh}\Phi_m}{8\pi\sigma r^3 \dot{\gamma} t N_A} \quad (30)$$

when $\frac{3\Delta E_{coh}\Phi_m}{8\pi\sigma r^3 \dot{\gamma} t N_A} = 1$, i.e. $\Delta E_{vap}/N_A = 8.37 \times 10^{-13} J$ under conditions $r$=100 μm and $\sigma\dot{\gamma}t = 0.072$ Pa, $\Phi_m$=0.72, then $\rho_{tap}/\rho_b$=1.2, quite close to the empirical criterion value, 1.22, as mentioned earlier. This point should be a critical turning point as it is obtained by assuming $\frac{d\eta}{d(\rho_t/\rho_b)} = 0$, implying that the viscosity would most likely have a minimum at this point. Strictly speaking, the minimum viscosity point should occur at the point where Hausner ratio is expressed with Eq. (30), not a fixed value and varying with systems; Only at a special case, $\frac{3\Delta E_{coh}\Phi_m}{8\pi\sigma r^3 \dot{\gamma} t N_A} = 1$, the minimum viscosity point occurs at $\rho_{tap}/\rho_b$=1.2. Eq. (24) together with Eq. (30) solves a long time controversial mystery in literature why Hausner ratio could be used as a flowability index for some materials but fails to work for others.

In a word, the viscosity equation of granular systems obtained with Eyring's rate process theory and Eyring's free volume estimation seems to show that viscosity may go through a minimum against the cohesive energy, the shear rate, the shear stress, and Hausner ratio. The competition between the shearing forces and the interparticle energy is believed to be responsible for this kind of phenomena. The empirical flowability criteria expressed with the popular Carr index and Hausner ratio may be reasoned with the obtained viscosity equation under a special circumstance. A long time confusion and controversy associated with Carr index and Hausner ratio for indicating powder flowability is thus cleared out with this new viscosity equation. One may be unable to numerically plot viscosity vs. particle size at same cohesive energy, as the cohesive energy typically changes with particle sizes. One important parameter missing in this viscosity equation is the particle volume fraction, which should play an important role in viscosity determination. Another parameter missing in this viscosity equation is the particle shape: spherical shape particles are definitely flow much better than needle-shape particles. However, this shape information should be indirectly included in bulk density data, as spherical shape particles tend to have higher bulk density than needle-shaped particles of same materials. Higher bulk densities mean lower Hausner ratios, thus lower viscosities before the critical points and higher viscosities after the critical points. A new approach to build up viscosity equations will be provided in next section for correlating the particle volume fraction information.

## 3. Viscosity equation derived from Eyring's rate theory and Hao's free volume estimation

In last section viscosity of granular materials is derived on the basis of Eyring's free volume equation, Eq. (14). In this section, Eyring's rate theory and Hao's free volume calculation will be used to derive viscosity equations of granular materials. The inter-particle spacing (IPS) that scales the distance between two particle surfaces would be used for estimating the free volume of whole system, as used by Hao for deriving the viscosity equations of colloidal suspension systems [41,42] and tap density equations of granular powders [50]. The free volume of an individual particle may be expressed as the space where an individual particle travels three dimensionally in a distance two times as long as the IPS [41,42,50]:

$$V_{fp} = (2IPS)^3 = 64(\sqrt[3]{\phi_m/\phi} - 1)^3 r^3 \qquad (31)$$

where $\phi_m$ is the maximum packing fraction that the particles can reach under a shear and $\phi$ is the particle volume fraction under current shear conditions. Theoretically, $\phi_m \rho_t = \rho_{tap}, \phi \rho_t = \rho_b$, therefore,

$$\frac{\phi_m}{\phi} = \frac{\rho_{tap}}{\rho_b} \qquad (32)$$

Eq. (31) represents the free volume of a single particle. The total free volume in a system should be [41,42]:

$$V_{tfp} = 64(\sqrt[3]{\phi_m/\phi} - 1)^3 r^3 \times \frac{\phi V_w}{4\pi r^3/3} = 15.29\phi(\sqrt[3]{\phi_m/\phi} - 1)^3 V_w \qquad (33)$$

where $V_w$ is the volume of the system. The free volume per unit volume of the whole system may be expressed as:

$$\bar{V}_f = V_{tfp}/V_w = 15.29\phi(\sqrt[3]{\phi_m/\phi} - 1)^3 \qquad (34)$$

If we still want to use Eq. (17) to derive the viscosity of granular materials with the free volume of particles per unit volume expressed in Eq. (34), one may need to modify Eq. (17), for the reason that the term $V_f$ in Eq. (17) may be considered as the energy needed to separate molecules per unit cell in the distance of $(V/N_A)^{1/3}$, rather than the total free volume available in the system. As mentioned earlier, the viscosity cannot be directly proportional to the free volume of the whole system. Instead, it should play in an opposite way, as the more the free volume is, the less the viscosity should be. When a granular system is under a shear stress σ and shear rate $\dot{\gamma}$, according to the literature [21], the shear force applied to the system per free volume, $\sigma_f$, over a time period of $t$ may be expressed as:

$$\sigma_f = \frac{\sigma \dot{\gamma} V t}{V_{tfp}} = \frac{\sigma \dot{\gamma} t}{\bar{V}_f} \tag{35}$$

where $V_{tfp}$ is the total free volume in the system expressed in Eq. (33) and $\bar{V}_f$ is the total free volume per unit volume and expressed in Eq. (34). So the energy needed for moving one particle per free volume unit in a distance of the inter-particle spacing may be written as:

$$E_f = \sigma_f \times 2IPS = \frac{4r\sigma \dot{\gamma} t}{15.29\phi(\sqrt[3]{\phi_m/\phi}-1)^2} \tag{36}$$

Replacing $V_f^{1/3}$ with $E_f$ expressed above in Eq. (17) yields

$$\eta = \frac{f}{2\Delta E} \bullet (2\pi m k_B T_{gp})^{1/2} \frac{4r\sigma \dot{\gamma} t}{15.29\phi} (\sqrt[3]{\phi_m/\phi}-1)^{-2} e^{E_0/k_B T_{gp}} \tag{37}$$

Again, taking the simple shear as an example, using $f/\Delta E = N_A/V$, $m = \frac{4}{3}\pi r^3 \rho_t$ and substituting $T_{gp}$ with Eq. (11) yields

$$\eta = \frac{N_A}{V} \bullet \frac{4\pi r^3 \rho_t}{3} (\frac{\pi \sigma \dot{\gamma} t}{3\rho_b})^{1/2} \frac{4r\sigma \dot{\gamma} t}{15.29\phi} (\sqrt[3]{\phi_m/\phi}-1)^{-2} e^{\frac{9E_0\rho_b}{8\pi\sigma \dot{\gamma} \rho_t t r^3}} \tag{38}$$

Note $V$ is the molar volume of particles expressed in Eq. (21). Substituting Eq. (21) for $V$ and Eq. (23) for $E_0$ into Eq. (38) yields

$$\eta = (\frac{\pi \sigma \dot{\gamma} t \rho_b}{3})^{1/2} \frac{4r\sigma \dot{\gamma} t}{15.29\phi} (\sqrt[3]{\phi_m/\phi}-1)^{-2} e^{\frac{3\Delta E_{vap}\rho_b}{8\pi\sigma \dot{\gamma} \rho_t t r^3 N_A}} \tag{39}$$

Note $\rho_b = \phi \rho_t$, replacing $\Delta E_{vap}$ with $\Delta E_{coh}$ and re-arranging Eq. (39) leads

$$\eta = 0.26 r (\frac{\pi \rho_t}{3})^{1/2} (\sigma \dot{\gamma} t)^{1.5} \phi^{1/6} (\phi_m^{1/3}-\phi^{1/3})^{-2} e^{\frac{3\Delta E_{coh}\phi}{8\pi\sigma \dot{\gamma} t r^3 N_A}} \tag{40}$$

Eq. (40) represents how viscosity changes with particle volume fraction, a very important parameter if a granular system is considered as particles dispersed in air, in analogy with colloidal suspensions where particles are dispersed into a liquid medium. Anything that may change particle volume fraction, like granulotemperature, shear stress, air pressure inside a powder system, interparticle forces, moisture, etc., could have an impact on viscosity. Note that Eq. (40) should be identical to Eq. (24), if the particle volume fraction in Eq. (40) is properly related to the granulotemperature. The big difference between Eq. (24) and Eq. (40) is that the viscosity in Eq. (24) is proportional to the quartic power of the particle radius, while it only has a linear relationship with the particle radius in Eq. (40); the cohesive energy per particle is included in Eq. (24) outside of the exponential term, however, it is missing in Eq. (40) besides that it appears in the exponential term. Keep in mind that the cohesive energy may be proportional to the molar volume of particles, thus both these two equations are essentially identical: it is linearly proportional to the particle radius if the cubic power of particle radius is taken away from the cohesive energy in the denominator of Eq. (24). Furthermore, Eq. (40) assumes that there is a maximum packing fraction for granular systems at which the interparticle spacing is zero and the whole system is completely jammed. In such a case the viscosity of whole system according to Eq. (40) turns to infinity. Figure 6 shows the predicted viscosity with Eq. (40) plotted against the particle volume fraction under a simple shearing with conditions $\sigma \dot{\gamma} = 1$ Pa, $\Delta E_{vap}/N_A = 5 \times 10^{-13}$ J, $r=100$ μm, particle true density, $\rho_t = 1000 \ kg/m^3$, and the random close packing is assumed with $\phi_m = 0.63$. The viscosity increases several orders of magnitudes when the particle volume fraction changes from 0.01 to 0.63. A dramatic increase of viscosity seems to start at the particle volume fraction about 0.4 and viscosity finally reaches to infinity once the particle volume fraction approaches to the maximum packing fraction, 0.63, at random close packing assumption. This is very similar to a percolation process where the physical properties of the whole system are dependent on how the system is percolated from one end to another due to the crowding effect. The infinity viscosity implies that the granular system is jammed and could not flow any more.

Eq. (40) shows that the viscosity apparently is dependent on the shear rate and shear stress, too, and the dependence should be very similar to what is shown in Fig. 4 and 5. Figure 7 shows the predicted viscosity against the shear rate under conditions $r=100$ μm, $\rho_t = 1000 kg/m^3$, $\Delta E_{coh}/N_A = 5 \times 10^{-13} J$, $\sigma \dot{\gamma}=1$ Pa•s, and $\phi_m = 0.63$, with two different particle volume fractions, Φ=0.2 and Φ=0.6, respectively. A similar shearing thinning phenomenon followed by shear thickening phenomenon is predicted. Again, when the viscosity times the shear rate is higher than the shear stress, the system starts to jam even when the particle volume fraction is much less than the maximum volume fraction. This conclusion seems qualitatively consistent with the computer simulation results that the jammed states may be diagramed with both shear stress and particle volume fraction [85]. The curves of viscosity vs. shear rate shown in Figure 7 are very similar to that of Figure 4, indeed, and a same trend will be observed for shear stress, too. Again, Figure 7 demonstrates that the shear thickening is dependent on shear rate, which is a puzzling issue in highly concentrated colloidal suspensions containing non-Brownian particles, as shear rate independence is predicted but untrue. Eq. (40) resolves the shear rate dependence issue, but again predicts a continuous shear thickening rather than discontinuous shear thickening. Intuitively, once the Brownian motion disappears in colloidal suspensions, those systems should behave identically like granular powders. In this scenario, particles will start to jam at a shear rate where the generated shear stress is larger than the applied shear stress, as indicted previously; A higher applied shear stress will force the jammed state to melt, until reach the next jamming state. The jamming-melting-jamming cycle will continue for a long time of period, creating a continuous shear thickening phenomena; The discontinuous shear thickening should never directly happened in this framework, however, have been observed in many highly concentrated colloidal suspensions. The discrepancy could be resulted from concurrences induced by shear that will be addressed by the end of this section, or from extrinsic factors during measurements such as the geometric confinement or space constraints [36,37,38].

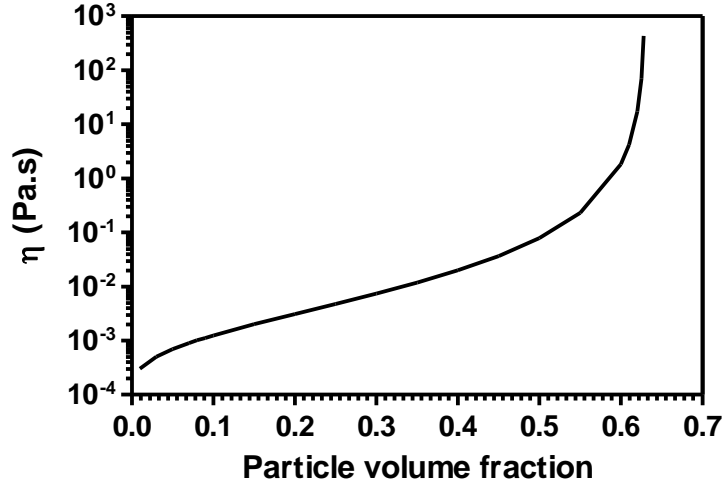

*Figure 6. The predicted viscosity with Eq. (40) vs. the particle volume fraction under a simple shearing $\dot{\sigma\gamma}t = 1$ Pa, $\Delta E_{coh}/N_A = 5 \times 10^{-13}$ J, r=100 μm and particle true density, $\rho_t = 1000$ kg/m³, The random close packing is assumed with $\phi_m = 0.63$.*

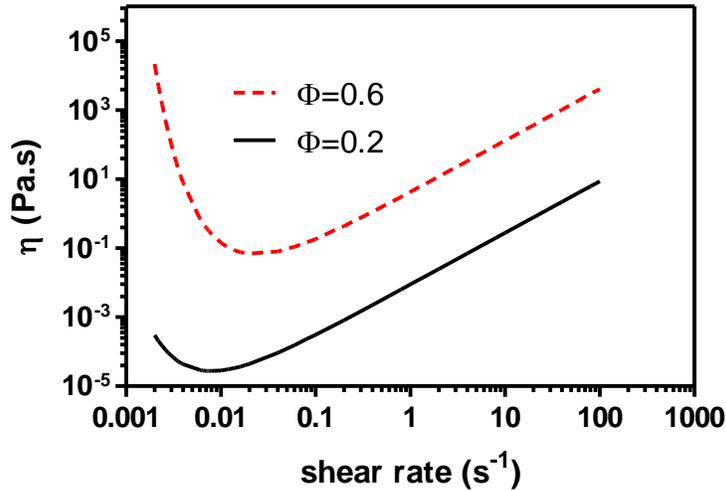

*Figure 7. The predicted viscosity with Eq. (40) vs. shear rate under conditions r=100 μm, $\rho_t$=1000 kg/m³, $\Delta E_{coh}/N_A$=5×10⁻¹³ J, $\sigma t$=1 Pa•s, and $\phi_m = 0.63$.*

The viscosity dependence on the cohesive energy predicted with Eq. (40) should be quite different than Eq. (24). Figure 8 shows the viscosity against the cohesive (or adhesive) energy between particles predicted with Eq. (40), under assumptions that particle radius $r=100$ $\mu m$, $\rho_t=1000$ kg/m$^3$, $\phi_m = 0.63$, $\sigma\dot{\gamma}t=1$, obtained when particle volume fractions are 0.2 and 0.6, respectively. The system of particle volume fraction 0.6 is predicted to have a higher viscosity than that of particle volume fraction 0.2, which intuitively makes sense. For both particle volume fractions, the viscosity doesn't change with the cohesive energy, until the cohesive energy reaches a critical point, about $10^{-10}$ J. Then the viscosity dramatically increases with the cohesive energy. This transition could be resulted from the particle jamming where the cohesive energy between particles is comparable or larger than the shear forces applied to the system and the system start to jam. Such a jamming transition even happens for particle volume fraction as low as 0.2. This type of viscosity jump is again very similar to the discontinuous shear thickening observed in many highly concentrated colloidal suspensions. For the purpose of seeing clearly how the shear forces will impact on the viscosity when the cohesive energy varies, the predicted viscosity with Eq. (40) is plotted against the cohesive energy under two shearing conditions in Figure 9. When the shear force indicated by $\sigma\dot{\gamma}t$ is as low as 0.1 $Pa$, the system shows a lower viscosity in comparison with an identical system at the high shear force condition, $\sigma\dot{\gamma}t=1$. With the increase of the cohesive energy, the viscosity of the system under a low shear force condition dramatically increases, again similar to the discontinuous shear thickening. In contrast, at a high shear force the viscosity only increases a little bit. Clearly, this is resulted from the competition between the shear force and the cohesive energy. If the applied shear force is larger than the cohesive energy between particles, the system yields and shows a low viscosity; otherwise, the system shows a very high viscosity, a very reasonable prediction.

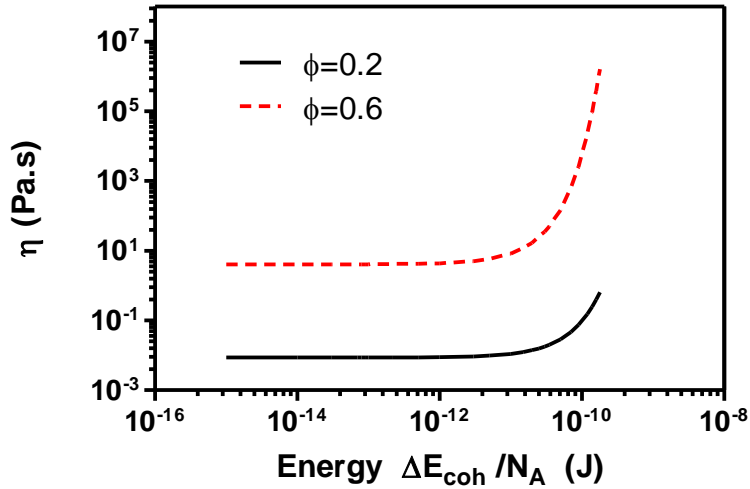

*Figure 8 The viscosity vs. the cohesive (or adhesive) energy between particles predicted with Eq. (40). Particle radius $r=100$ $\mu m$, $\rho_t=1000$ kg/m$^3$, $\phi_m = 0.63$, $\sigma\dot{\gamma}t=1$.*

It is worth noting that although Eq. (40) doesn't directly predict a discontinuous shear thickening behaviors as shown in Figure 7, Figures 6, 8 and 9 show that the discontinuous thickening like behaviors may still happen, when the particle volume fractions approach to the maximum packing fraction, when the cohesive energy between particles exceeds certain values, and when the applied shear energy ($\sigma \dot{\gamma} t$) is small enough. If any those three conditions are met during the increase of shear rates, the discontinuous shear thickening phenomenon will be observed. Recent numerical simulation work [35,39] indicates that the discontinuous shear thickening phenomenon will be generated once the interparticle frictional coefficient is larger than zero, supporting the derivation from Eq. (40) that the large enough cohesive energy will lead to abrupt viscosity increase, as the cohesive energy between particles could simply come from the frictional interactions, electrostatic attractions, or other forces. The discontinuous shear thickening studied in cornstarch/water suspensions using a rheometer coupled with local Magnetic Resonance Imaging (MRI) measurements reveals that the discontinuous shear thickening only occurs when an inhomogeneity is induced during shear and the suspension is separated into a low-density unjammed and a high-density jammed region [86]. Again, the high-density jammed region means that the particle volume fraction in this region could very likely approach to the maximum packing fraction, consistent with what is demonstrated in Figure 6. Two of three conditions proposed for discontinuous shear thickening are evidenced experimentally, and the third condition related to the applied shear energy could be easily understood, as the shear thinning and thickening should be determined by the competition between the applied shear energy and the cohesive energy between particles: a weaker shear energy in comparison with the cohesive energy between particles will lead to the shear thickening, otherwise shear thinning.

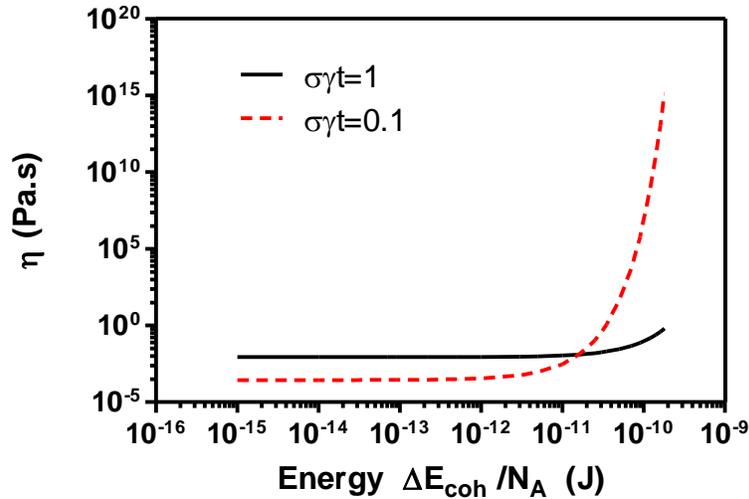

*Figure 9 The viscosity vs. the cohesive (or adhesive) energy between particles predicted with Eq. (40). Particle radius $r=100$ μm, $\rho_t=1000$ $kg/m^3$, $\phi_m = 0.63$, $\phi=0.2$.*

## III Discussion

The viscosity concept of granular powders looks really odd at the beginning, as granular powder systems are athermal systems and there is no molecular level frictional forces resulted from the movements of particles. However, under the proper definition of granular temperature, granular powders will gain an analogous and meaningful flowability indicator as usually used in thermal systems, the viscosity. Defining the viscosity in an uniformed manner and deriving viscosity equations using the same uniformed approaches that have been proved to work for the liquids, colloidal suspensions, and polymeric systems, could provide an easy and comprehensible means to understand granular powder systems. Eyring's rate process theory and free volume concept are two very powerful tools widely employed to treat many problems in physical chemistry, physics, and material sciences. They are borrowed again to treat granular powders and the obtained viscosity equations definitely provide many insightful clues on powder flowability. Together with my early publications employing the rate process theory and the free volume concept for deriving viscosity equations of pure liquids, colloidal suspensions, and polymeric melts and solutions [41,42], both thermal and athermal systems are therefore unified with a single approach.

The approach presented in this article distinguished from others lies on analogously defining viscosity of dry granular powders after granular temperature is defined consistently with the traditional temperature. However, this is my first attempt and it wouldn't resolve every issue in dry granular powders area. Since there is no a systematical way to experimentally measure the viscosity of granular powders as we usually do for liquids, colloidal suspensions, and solids, the obtained viscosity equations may only be compared with empirical macroscopic observations of dry granular powders, which are demonstrated across my article whenever possible. An example of those successes is that the equation (24) provides a theoretical basis for widely used empirical powder flowability indicators in many industrial areas, Hausner ratio or Carr index. Another successful example is that the predicted shear thickening from my viscosity equations, Eq. (24) and (40), is dependent on shear rate, the cohesive energy between particles, the particle volume fractions, as demonstrated from Figures 3 to 9. Both Eq. (24) and (40) predict a continuous shear thickening rather than the discontinuous shear thickening, which is reasonable as the jamming-melting-jamming cycles should always happen in granular powders. Nonetheless, the discontinuous shear thickening may still happen, if one of the three conditions below could be induced and met during shear: the particle volume fractions approach to the maximum packing fraction; the cohesive energy between particles exceeds certain values; and the applied shear energy ($\sigma \dot{\gamma} t$) is small enough. It is those concurrences induced by shear that indirectly contribute to the discontinuous shear thickening.

From viscosity equations and all figures presented in this article, it looks like the fundamental mechanisms of how dry granular particles flow may be concisely described with the following scenario. If the applied shear energy even is slightly larger than the cohesive energy between particles, particles start to flow and an immediate shear thinning phenomena should be always observed due to a cascade of structural unbalances in powder beds. Once the applied shear field is strong enough capable of driving particles to move, the induced particle-particle interaction may create resistance to the flow and thus the shear thickening starts to appear. Initially the generated resistance should be smaller than the applied shear force, the system should freely flow. Further increase of shear rates may make the resistance force larger than the applied shear force, the system starts to jam. After this the system enters jamming-melting-jamming cycles as described earlier. In such a scenario, the continuous shear thickening is therefore expected, as jamming-melting-jamming cycles may last a long period of time. During vigorous shear, particle segregation [87,88] may occur, creating inhomogeneities in the system where some regions have more particles than others; If the particle volume fractions in these regions are high enough that the particle interaction forces or called the cohesive energy between particles are much stronger than the applied shear forces, or the particle volume fraction approaches to the maximum packing fractions and a strong locking effect is generated due to the geometric constraints, the system is jammed and the discontinuous shear thickening thus occurs. It looks like that the discontinuous shear thickening is the indirect effect induced by the shear rate.

**IV Summaries and conclusions**

Using the analogous temperature defined in granular powders, the viscosity equations of granular powders are derived on the basis of Eyring's rate process theory and free volume concept. The viscosity equations are obtained with three different ways: 1) derived with Eyring's rate process theory only; 2) derived with Eyring's rate process theory and Eyring's free volume method; 3) derived with Eyring's rate process theory and Hao's free volume calculation approach. The first approach leads to a similar Arrhenius viscosity equation with unknown parameter *A*. The second approach leads to a very successful viscosity equation, showing that viscosity may go through a minimum against the cohesive energy, the shear rate, the shear stress, and Hausner ratio. It also can clearly tell why the Carr index or the Hausner ratio can be used to empirically scale powder flowability and provide a theoretical basis for the empirical criteria widely used in many industrial fields. The third approach correlate viscosity with particle volume fractions and clearly demonstrate the viscosity of granular powders is strongly dependent on the competition between the shear force and the cohesive energy between particles.

The obtained viscosity equations show that the viscosity of granular powders are complicatedly dependent on powder densities both true and bulk densities, the applied shear stress and shear rate, the shear time, the cohesive energy between particles, the particle volume fractions, and the particle size. Although the derived viscosity equations, Eq. (24) and (40), predict a continuous shear thickening rather than the discontinuous shear thickening, they clearly tell that the discontinuous shear thickening may still occur when one of the following three cases are induced and met during shear: the cohesive energy between particles exceeds a critical value; the applied shear force is small enough in comparison with the cohesive energy; and the particle volume fraction approaches to the maximum packing fraction. It is those concurrences induced by shear that indirectly contribute to the discontinuous shear thickening.

A fundamental mechanism on how dry granular powders flow may be described below: particles may easily flow even under a small shear due to the weak structural nature of powder systems. A cascade of unbalanced structural failures due to particles moving of supportive positions leads to an initial shear thinning phenomena. Continuous increase shear energy brings granular powders into free flowing regime, until the generated shear resistance is larger than the applied shear force and the systems start to jam. After that the systems enter jamming-melting-jamming cycles, displaying the continuous shear thickening or discontinuous shear thickening, dependent on if local particle volume fractions or the interparticle forces are high enough to cause systematical jamming. All predictions are consistent with experimental evidences, empirical observations, and general intuitions. The current work may lay a foundation for quantitatively scaling the flowability of granular powder, thus has a significant impact on industrial applications, too.

# References

1. P. M. Reis, R. A. Ingale, and M. D. Shattuck, Caging Dynamics in a Granular Fluid. *Phys. Rev. Lett.* **98**, 1883011-4 (2007).
2. P. M. Reis, R. A. Ingale, and M. D. Shattuck, Crystallization of a Quasi-Two-Dimensional Granular Fluid. *Phys. Rev. Lett.* **96**, 2580011-4 (2006).
3. S. Savage and D. Jeffery, The stress tensor in a granular flow at high shear rates. *J. Fluid Mech.* **110**, 255-272 (1981).
4. C. Lun, S. B. Savage , D. J. Jeffrey, and N. Chepurniy, Kinetic theories for granular flow: inelastic particles in Couette flow and slightly inelastic particles in a general flowfield. *J. Fluid Mech.* **140**, 223-256 (1984).
5. C. Lun, Kinetic theory for granular flow of dense, slightly inelastic, slightly rough spheres. *J. Fluid Mech.* **233**, 539-559 (1991).
6. Shapiro, A. G. a. M., Mechanics of collisional motion of granular materials. Part 1. General hydrodynamic equations. *J. Fluid Mech.* **282**, 75-114 (1995).
7. N. Sela, I. Goldhirsch, and S. H. Noskowicz, Kinetic theoretical study of a simply sheared two-dimensional granular gas to Burnett order. *Phys. Fluids* **8**, 2337-2353 (1996).
8. J. J. Brey, F. Moreno, and J. W. Dufty, Model kinetic equation for low-density granular flow. *Phys. Rev. E* **54**, 445-456 (1996).
9. J. J. Brey and J. W. Dufty, Hydrodynamic modes for a granular gas from kinetic theory. *Phys. Rev. E* **72**, 011303-16 (2005).
10. J. F. Lutsko, Chapman-Enskog expansion about nonequilibrium states with application to the sheared granular fluids. *Phys. Rev. E* **73**, 021302-19 (2006).
11. L. B. Loeb, *The Kinetic Theory of Gases* (Dover, New York, 2004).
12. A. Santos, J. M. Montanero, J. W. Dufty, and J. J. Brey, Kinetic model for the hard-sphere fluid and solid. *Phys.Rev. E* **57**, 1644-1660 (1998).
13. V. Garzó and J. Dufty, Homogeneous cooling state for a granular mixture. *Phys. Rev. E* **60**, 5706-5713 (1999).
14. J. W. Dufty, A. Baskaran, and L. Zogaib, Gaussian kinetic model for granular gases. *Phys. Rev. E* **69**, 051301-21 (2004).
15. V. Kumaran, Kinetic Model for Sheared Granular Flows in the High Knudsen Number Limit. *Phys. Rev. Lett.* **95**, 108001-4 (2005).
16. J. Jenkins and M. Richman, Kinetic theory for plane flows of a dense gas of identical, rough, inelastic, circular disks. *Phys. Fluids* **28**, 3485-3494 (1985).
17. H. M. Jaeger, Celebrating Soft Matter's 10th Anniversary: Toward jamming by design. *Soft Matter* **11**, 12-27 (2015).
18. D. Bi, J. Zhang, B. Chakraborty, and R. P. Behringer, Jamming by shear. *Nature* **480**, 355-358 (2011).
19. V. Trappe, V. Prasad, L. Cipelletti, P. N. Segre, and D. A. Weitz, Jamming phase diagram for attractive particles. *Nature* **411**, 772-775 (2001).
20. Z. Zhang, N. Xu, D. T. N. Chen, P. Yunker, A. M. Alsayed, K. B. Aptowicz, P. Habdas, A. J. Liu, S. R. Nagel, and A. G. Yodh, Thermal vestige of the zero-temperature jamming

transition. *Nature* **459**, 230-233 (2009).

21. T. Hao, Defining Temperatures of Granular Powders Analogously with Thermodynamics to Understand the Jamming Phenomena. *submitted* (2015).
22. P. Juliano, and G. V. Barbosa-Cánovas, Food Powders Flowability Characterization: Theory, Methods, and Applications. *Ann. Rev. Food Sci.Tech.* **1**, 211-239 (2010).
23. V. Ganesana, K. A. Rosentrater, K. Muthukumarappan, Flowability and handling characteristics of bulk solids and powders – a review with implications for DDGS. *Biosys. eng.* **101**, 425–435 (2008).
24. K. Hutter, Geophysical granular and particle-laden flows:review of the field. *Phil. Trans. R. Soc. A* **363**, 1497–1505 (2005).
25. Y. Forterre and O. Pouliquen, Flows of dense granular media. *Annu. Rev. Fluid Mech.* **40**, 1-24 (2008).
26. C. S. Campbell, Rapid Granular Flows. *Ann. Rev. Fluid Mech.* **22**, 57-90 (1990).
27. P. Jop, Y. Forterre, O. Pouliquen, A constitutive law for dense granular flows. *Nature* **441**, 727-730 (2006).
28. F. Boyer, E. Guazzelli, O. Pouliquen, Unifying suspension and granular rheology. *Phys. Rev. Lett.* **107**, 188301 (2011).
29. R. S. Saksenaa and L. V. Woodcock, Quasi-thermodynamics of powders and granular dynamics. *Phys. Chem. Chem. Phys.* **6**, 5195-5202 (2004).
30. M. P. Ciamarra, A. Coniglio, and M. Nicodemi, Thermodynamics and Statistical Mechanics of Dense Granular Media. *Phys. Rev. Lett.* **97**, 158001 (2006).
31. J. Casas-Vazquez, and D. Jou, Temperature in non-equilibrium states: a review of open problems and current proposals. *Rep. Prog. Phys.* **66**, 1937-2023 (2003).
32. K. Lu, E. E. Brodsky, and H. P. Kavehpour, A thermodynamic unification of jamming. *Nature Phys.* **4**, 404-407 (2008).
33. Q. Chen, and M. Hou, Effective temperature and fluctuation-dissipation theorem in athermal granular systems: A review. *Chin.Phys. B* **23**, 074501 (2014).
34. M. Leturia, M. Benali, S. Lagarde, I. Ronga, and K. Saleh, Characterization of flow properties of cohesive powders: A comparative study of traditional and new testing methods. *Powder Tech.* **253**, 406–423 (2014).
35. R. Seto, R. Mari, J. F. Morris,and M. M. Denn, Discontinuous Shear Thickening of Frictional Hard-Sphere Suspensions. *Phys. Rev. Lett.* **111**, 218301 (2013).
36. Q. Xu, S. Majumdar, E. Brown and H. M. Jaeger, Shear thickening in highly viscous granular suspensions. *Europhys. Lett.* **107**, 68004 (2014).
37. X. Bian, S. Litvinov, M. Ellero, and N. J. Wagner, Hydrodynamic shear thickening of particulate suspension under confinement. *J. Non-Newtonian Fluid Mech.* **213**, 39-49 (2014).
38. E. Brown and H. M. Jaeger, Shear thickening in concentrated suspensions: phenomenology, mechanisms, and relations to jamming. *Rep. Prog. Phys.* **77**, 046602 (2014).
39. R. Mari, R. Seto, J. F. Morris, and M. M. Denn, Shear thickening, frictionless and frictional rheologies in non-Brownian suspensions. *J. Rheol.* **58**, 1693-1724 (2014).
40. M. Wyart and M. Cates, Discontinuous shear thickening without inertia in dense non-Brownian suspensions. *Phys. Rev. Lett.* **112**, 098302 (2014).

41. T. Hao, *Electrorheological Fluids: The Non-aqueous Suspensions* (Elsevier Science, Amsterdam, 2005).
42. T. Hao, Viscosities of liquids, colloidal suspensions, and polymeric systems under zero or non-zero electric field. *Adv. Coll. Interf. Sci.* **142**, 1-19 (2008).
43. S. Glasstone, K. Laidler, and H. Eyring, *The theory of rate process* (McGraw-Hill, New York, 1941).
44. A. J. Kovacs, Applicability of the free volume concept on relaxation phenomena in the glass transition range. *Rheol. Acta* **5**, 262-269 (1966).
45. V. S. Nechitailo, About the Polymer Free Volume Theory. *Intern. J. Polym. Mat.* **16**, 171-177 (1992).
46. H. Eyring and J. O. Hirschielder, The theory of the liquid state. *J. Phys. Chem.* **41**, 249-257 (1937).
47. P. Hänggi, P. Talkner, and M. Borkovec, Reaction-rate theory: fifty years after Kramers. *Rev. Mod. Phys.* **62**, 251-341 (1990).
48. F. H. Stillinger, Z. W. Salsburg, and R. L. Kornegay, Rigid Disks at High Density. *J. Chem. Phys.* **43**, 932-943 (1965).
49. W. G. Hoover, N. E. Hoover, and K. Hansonc, Exact hard-disk free volumes. *J. Chem. Phys.* **70**, 1837-1844 (1979).
50. T. Hao, Tap density equations of granular powders based on the rate process theory and the free volume concept. *Soft Matter* **11**, 1554-1561 (2015).
51. T. Hao, Derivation of stretched exponential tap density equations of granular powders. *Soft Matter* **11**, 3056-3061 (2015).
52. H. Eyring, Viscosity, Plasticity, and Diffusion as Examples of Absolute Reaction Rates. *J. Chem. Phys* **4**, 283-291 (1936).
53. S. A. Arrhenius, The viscosity of solutions. *Biochem. J.* **11**, 112–133 (1917).
54. Q. Li, V. Rudolph, B. Weigl, and A. Earl, Interparticle van der Waals force in powder flowability and compactibility. *Int. J. Pharm.* **280**, 77–93 (2004).
55. K. M. Beazley, in *Rheometry: Industrial Applications*, edited by Walters, K. (John Wiley & Sons, 1980), pp. 339-413.
56. L. B. Chen, B. J. Ackerson, and C. F. Zukoski, Rheological consequences of microstructural transition in colloidal crystals. *J. Rheol.* **38**, 193-216 (1994).
57. H. A. Barnes, J. F. Hutton, and K. Walters, *An introduction to Rheology, Amsterdam: Elsevier* (Elsevier, Amsterdam, 1989).
58. M. K. Chow, and C. F. Zukoski, Gap size and shear history dependencies in shear thickening of a suspension ordered at rest. *J. Rheol.* **39**, 15-32 (1995).
59. R. G. Larson, *The structure and rheology of complex fluids* (Oxford University Press, New York, 1999).
60. F. Ianni, D. Lasne, R. Sarcia, and P. Hébraud, Relaxation of jammed colloidal suspensions after shear cessation. *Phys. Rev. E.* **74**, 0114011-0114016 (2006).
61. E. Bertrand, J. Bibette, and V. Schmitt, From shear thickening to shear-induced jamming. *Phys. Rev. E.* **66**, 060401(R)1-3 (2002).
62. K. R. Stratford, K., R. Adhikari, I.Pagonabarraga, J.-C. Desplat, and M. E. Cates, Colloidal

Jamming at Interfaces: A Route to Fluid- Bicontinuous Gels. *Science* **309**, 2198-2201 (2005).
63. P. Varadan, and M. J. Solomon, Direct visualization of flow-induced microstructure in dense colloidal gels by confocal laser scanning microscopy. *J. Rheol.* **47**, 943-968 (2003).
64. E. Corwin, H. M. Jaeger, and S. R. Nagel, Structural signature of jamming in granular media. *Nature* **435**, 1075-1078 (2005).
65. P. Olsson, and S. Teitel, Critical Scaling of Shear Viscosity at the Jamming Transition. *Phys. Rev. Lett.* **99**, 1780011-4 (2007).
66. L. E. Silbert, D. Ertas¸ G. S. Grest, T. C. Halsey, and D. Levine, Analogies between granular jamming and the liquid-glass transition. *Phys. Rev. E* **65**, 0513071-4 (2002).
67. D. A. Weitz, Unjamming a Polymer Glass. *Science* **323**, 214-215 (2009).
68. H. Lee, K. Paeng, S. F. Swallen, M. D. Ediger, Direct Measurement of Molecular Mobility in Actively Deformed Polymer Glasses. *Science* **323**, 231-234 (2009).
69. R. K. Gupta, *Polymer and composite rheology* (Marcel Dekker, New York, 2000).
70. Q. D. Nguyen, and D. V. Boger, Measuring the flow properties of yield stress fluids, Annu. Rev. Fluid Mech. *Annu. Rev. Fluid Mech.* **24**, 47-88 (1992).
71. E. C. Bingham, *Fluidity and Plasticity* (McGraw-Hill, New York, 1922).
72. W. H. Herschel, and R. Bulkley, Measurement of consistency as applied to rubberbenzene solutions. *Proc. Am. Soc. Test. Matis.* **26**, 621-33 (1926).
73. W. Casson, in *Rheology of Dispersed Systems*, edited by Mill, C. C. (Pergamon Charm, S. E., London, 1959), pp. 84-104.
74. A. DeSimone, and C. Tamagnini, Stress–dilatancy based modelling of granular materials and extensions to soils with crushable grains. *Int. J. Numer. Anal. Meth. Geomech.* **29**, 73–101 (2005).
75. M. Cecconi, and G. M. B.Viggiani, Structural features and mechanical behaviour of a pyroclastic weak rock. *Int. J. Numer. Anal. Meth. Geomech.* **25**, 1525–1557 (2001).
76. O. Reynolds, On the dilatancy of media composed of rigid particles in contact, with experimental illustrations. *Phil. Mag., Series 5* **20**, 469–481 (1885).
77. O. Reynolds, Experiments showing dilatancy, a property of granular material, possibly connected with gravitation. *Proc. Royal Institution of Great Britain* **11**, 354-363 (1886).
78. T. Hao, Theoretical Bases of Hausner Ratio or Carr Index for Empirical Scaling the flowability of Granular Powders. *Submitted* (2015).
79. E. Guerin, P. Tchoreloff, B. Leclerc, D. Tanguy, M. Deleuil, G. Couarraze, Rheological characterization of pharmaceutical powders using tap testing, shear cell and mercury porosimeter. *Int. J. Pharm.* **189**, 91–103 (1999).
80. R. L. Carr, Evaluating flow properties of solids. *Chem. Eng.,* **72**, 163–168 (1965).
81. H. H. Hausner, Friction conditions in a mass of metal powder. *Int J Powder Metall.* **3**, 7–13 (1967).
82. M. De Villiers, in *Theory and Practice of Contemporary Pharmaceutics*, edited by T. K. Ghosh, B. R. J. (CRC Press, Boca Raton, 2005), pp. 298–299.
83. P. C. Seville, T. P. Learoyd, H.-Y. Li, I.J. Williamson, and J. C. Birchall, Amino acid-modified spray-dried powders with enhanced aerosolisation properties for pulmonary drug delivery. *Powder Tech.* **178**, 40-50 (2007).

84. A. Santomaso, P. Lazzaro, P. Canu, Powder flowability and density ratios: the impact of granules packing, Chem. Eng. Sci. *Chem. Eng. Sci.* **58**, 2857 – 2874 (2003).

85. M. P. Ciamarra, M. Nicodemi, and A. Coniglio, in *Proc. 6th intern. conf. on micromechan of granular media, Ed. M. Nakagawa and S. Luding* (American. Inst. Phys, 2009).

86. A. Fall, F. Bertrand, D. Hautemayou, C. Mezière, P. Moucheront, A. Lemaître, and G. Ovarlez, Macroscopic Discontinuous Shear Thickening versus Local Shear Jamming in Cornstarch. *Phys. Rev. Lett.* **114**, 098301 (2015).

87. J. M. N. T. Gray, and B. P. Kokelaar, Large particle segregation, transport and accumulation in granular free-surface flows. *J. Fluid Mech.* **652**, 105–137 (2010).

88. J. M. Ottino and D. V. Khakhar, Mixing and segregation of granular materials. *Annu. Rev. Fluid Mech.* **32**, 55–91 (2000).